\documentclass[%
 aip,
 amsmath,amssymb,
 reprint,%
]{revtex4-1}

\usepackage{graphicx}% Include figure files
\usepackage{dcolumn}% Align table columns on decimal point
\usepackage{bm}% bold math
\usepackage{xcolor}
\usepackage[utf8]{inputenc}
\usepackage[T1]{fontenc}
\usepackage{mathptmx}
\usepackage{etoolbox}
\usepackage{upgreek}
\usepackage{comment}

\def\cm{cm$^{-1}$}

\draft % marks overfull lines with a black rule on the right

\makeatletter
\def\@email#1#2{%
 \endgroup
 \patchcmd{\titleblock@produce}
  {\frontmatter@RRAPformat}
  {\frontmatter@RRAPformat{\produce@RRAP{*#1\href{mailto:#2}{#2}}}\frontmatter@RRAPformat}
  {}{}
}%
\makeatother

\begin{document}

%\preprint{}

\title[Raman Scattering Spectra of Boron Imidazolate Frameworks Containing Different Magnetic Ions]{Raman Scattering Spectra of Boron Imidazolate Frameworks Containing Different Magnetic Ions} %Title of paper



\author{Jackson Davis}
\affiliation{Department of Physics and Astronomy, Johns Hopkins University, Baltimore, Maryland 21218, USA}
\author{Soumyodip Banerjee}
\affiliation{Department of Chemistry, Johns Hopkins University, Baltimore, Maryland 21218, USA}
\author{Pilar Beccar-Varela}
\affiliation{Department of Chemistry, Johns Hopkins University, Baltimore, Maryland 21218, USA}
\author{V. Sara Thoi}
\affiliation{Department of Chemistry, Johns Hopkins University, Baltimore, Maryland 21218, USA}
\affiliation{Department of Materials Science and Engineering, Johns Hopkins University, Baltimore, Maryland 21218, USA}
\author{Natalia Drichko}
\email{drichko@jhu.edu}
\affiliation{Department of Physics and Astronomy, Johns Hopkins University, Baltimore, Maryland 21218, USA}

%\author{}
%\thanks{}
%\affiliation{}


\date{\today}

\begin{abstract}
We present a Raman scattering spectroscopic study of boron imidazolate  metal-organic frameworks (BIFs) with three different magnetic metal ions and one non-magnetic in a wide frequency range from 25 to 1700~\cm, which covers local vibrations of the linkers and well as collective lattice vibrations. We show that the spectral region above 800~\cm~belongs to the local vibrations of the linkers, which have the same frequencies for the studied BIFs without any dependence on the structure of the BIFs, and are easily interpreted based on the spectra of imidazolate linkers. In contrast, collective lattice vibrations, observed below 100~\cm, show a distinction between cage and two-dimensional  BIFs structures, with a weak dependence on the metal node. We identify the range of vibrations around 200~\cm, which are distinct for each MOF, depending on a metal node. Our work demonstrates the energy hierarchy in the vibrational response of BIFs.    
\end{abstract}


\maketitle %\maketitle must follow title, authors, abstract and \pacs

% References should be done using the \supercite, \ref, and \label commands

\section{Introduction}
Much of the study of metal-organic frameworks (MOFs) in recent years has focused on their tunability and porosity. Since MOF structures can include magnetic ions or clusters connected by organic linkers, the possibility of achieving interesting magnetic states in  MOFs has been discussed~\cite{Thorarinsdottir2020, MinguezEspallargas2018}. However, most of the recent studies of magnetic properties in MOFs have been limited by basic magnetization and magnetic susceptibility measurements. In order to study magnetism in these materials we will need to employ spectroscopy, such as magnetic Raman scattering. In fact, magnetic Raman scattering has demonstrated its ability to probe the spectrum of magnetic excitations in organic magnetic materials~\cite{Drichko2015,Hassan2018} where magnetic neutron scattering is challenging due to weak signals and presence of hydrogen in the materials. In order to use Raman scattering spectroscopy as a tool to study different MOF structures and their magnetic response, we need to obtain information about the energy scales of all the other  excitations; we must identify what part of the acquired spectroscopic information is related to the collective lattice modes, and what part is related to the linker vibrations. Vibrational Raman scattering has been widely used as a characterization tool for MOFs, as the vibrational modes within the organic ligands, and between the ligands and the metal ions, tend to produce strong peaks in the Raman spectra~\cite{Hadjiivanov2021}. In this manuscript, we present our vibrational Raman scattering studies of a range of MOFs which share the same linkers, but include different transitional metal ions and show three different structures. We aim to understand the hierarchy of the energy scales we observe in the very rich spectra of MOFs, and develop an efficient way to interpret the spectral features. 

\begin{figure}
    \includegraphics[width=\linewidth]{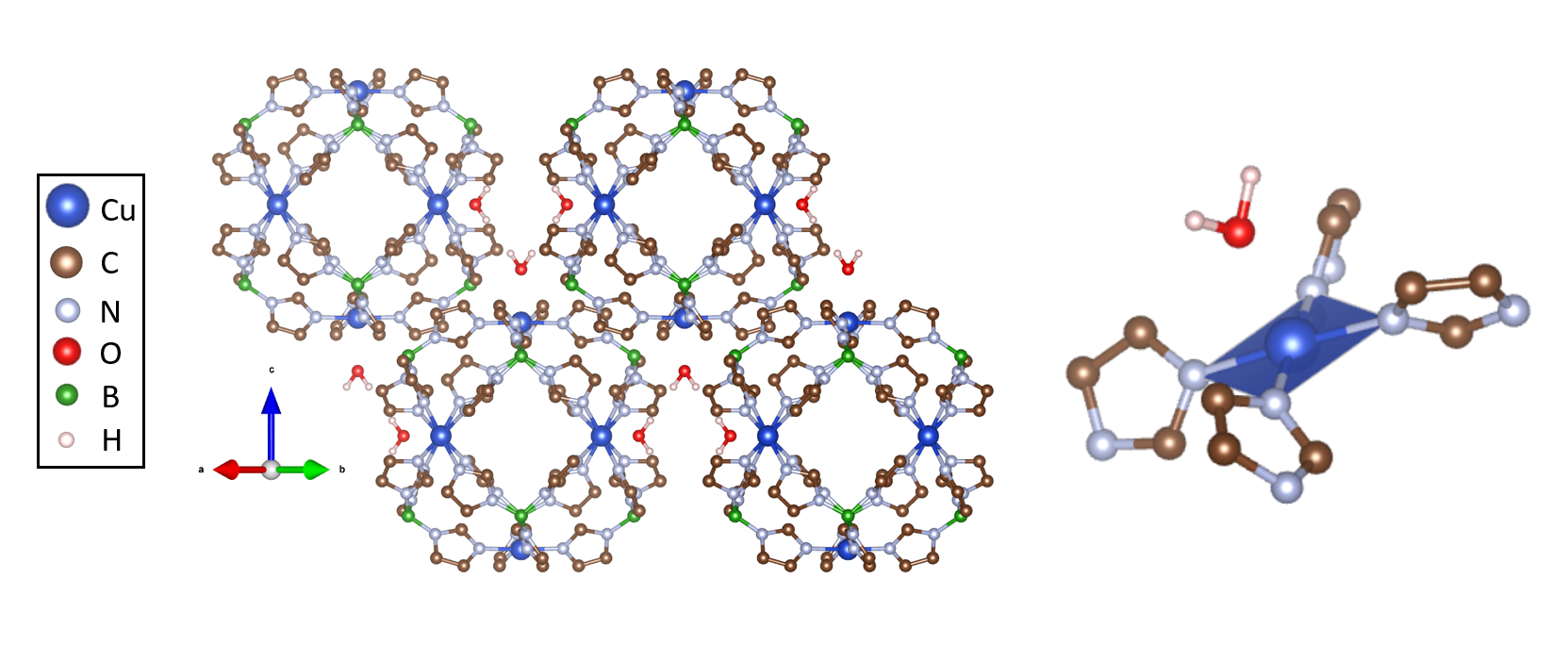}
    \caption{Left panel: Cu-BIF cage structure~\cite{Banerjee2022a}. Note the 4-membered ring structure of the pore. Right panel: anisotropic square pyramidal Cu environment. H atoms omitted for clarity, except those on H$_2$O that bind to Cu. All structural models visualized in VESTA~\cite{Momma2011}.}
    \label{fig:Cu_BIF}
\end{figure}

\begin{figure}
    \includegraphics[width=\linewidth]{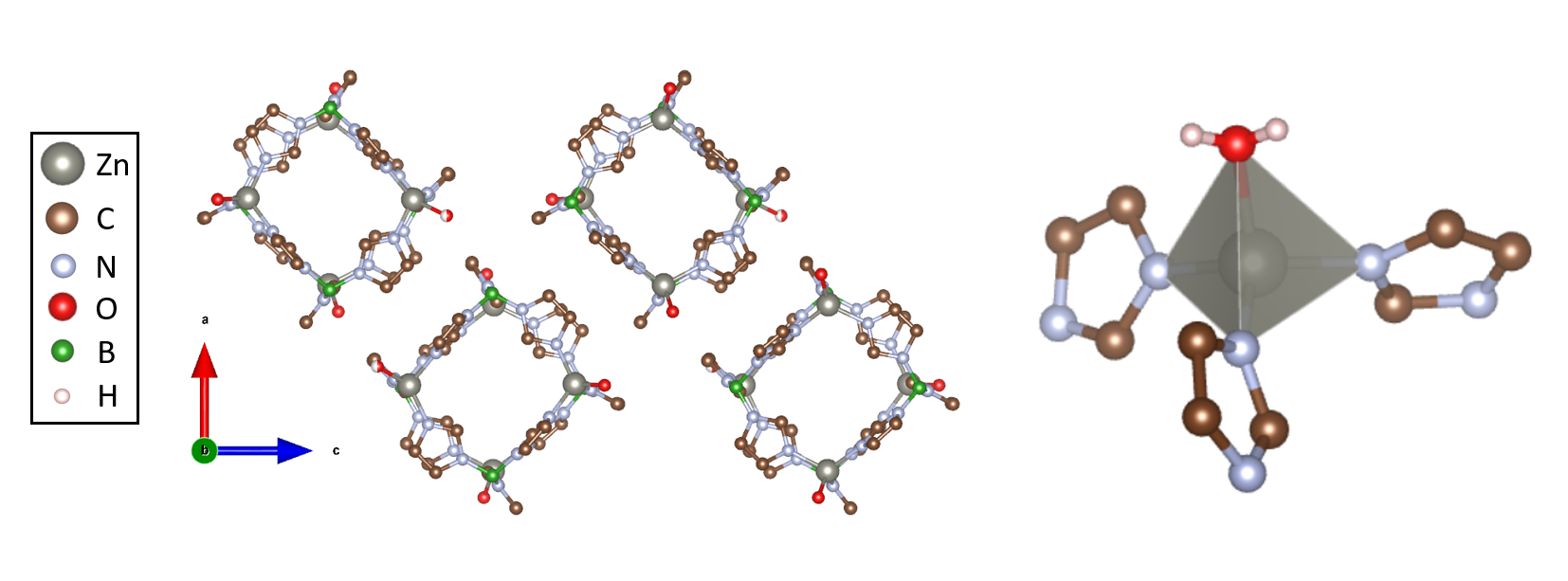}
    \caption{Left panel: Zn-BIF cage structure~\cite{Wen2017}. Note the preserved 4-membered ring structure of the pore, while the structure of the cage is different from  Cu-BIF. Right panel: anisotropic tetrahedral Zn environment. H atoms omitted for clarity, except those on H$_2$O that bind to Zn.}
    \label{fig:Zn_BIF}
\end{figure}

All MOF materials studied here belong to the class of boron imidazolate frameworks (BIFs), which consist of metal ions coordinated by a boron imidazolate ligand.  We present Raman spectra of four different BIFs with magnetic ions Cu$^{2+}$ (S= 1/2), Ni$^{2+}$ (S=1), Co$^{2+}$ (S= 3/2) and non-magnetic  Zn$^{2+}$  as the metal cations. These BIFs have three distinct structures: Cu-BIF and Zn-BIF each possesses one of two cage-like structures~\cite{Zhang2015, Wen2017}, while Co-BIF and Ni-BIF are isostructural, possessing a layered 2D triangular lattice~\cite{Banerjee2022}. The cage BIFs have topologies similar to zeolites and zeolitic imidazolate frameworks (ZIFs). In ZIFs, each metal cation bonds to one nitrogen of four different imidazolate rings, creating a tetrahedral metal environment, and these linked tetrahedra form a porous 3D network of connected cages~\cite{Chen2014}. In cage BIFs, one N on the imidazolate ring binds to a metal cation, while the other binds to a B$^{3+}$ cation. For instance, in the cage Cu-BIF, pictured in Fig.~\ref{fig:Cu_BIF}, the metal cations have an environment of four coplanar N atoms from four different imidazolate rings, while one oxygen from one H$_2$O weakly binds to the metal in the out-of-plane direction, creating a low symmetry environment for the metal atom compared to standard inorganic oxides. The Zn-BIF, pictured in Fig.~\ref{fig:Zn_BIF}, is composed of smaller cages of 4 Zn and 4 boron imidazolate ligands, and exhibits an anisotropic tetrahedral Zn environment.

\begin{figure}
    \includegraphics[width=\linewidth]{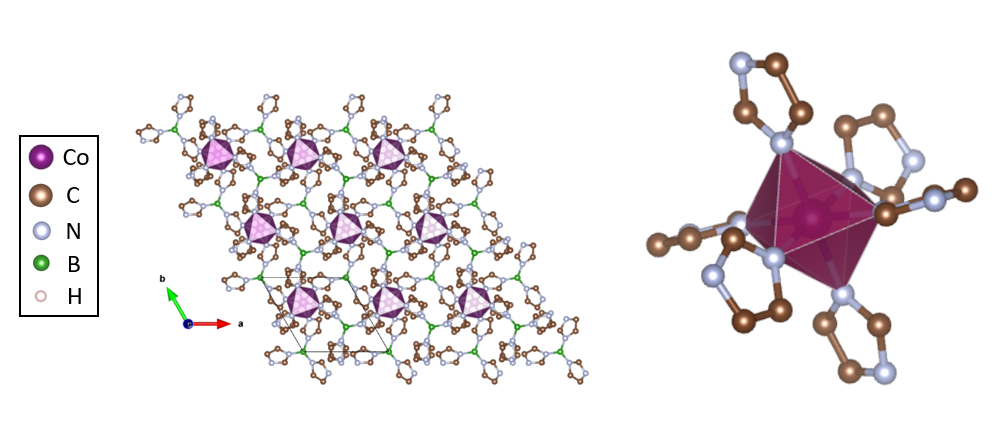}
    \caption{Left panel: Triangular lattice structure of Co-BIF in (ab) layer~\cite{Banerjee2022}. Right panel: octahedral Co environment. H atoms omitted for clarity. Ni-BIF is isostructural to Co-BIF~\cite{Banerjee2022}}
    \label{fig:Co_BIF}
\end{figure}

In 2D layered BIFs, no cages are formed at all, and metal ions arrange in a 2D triangular lattice (see Co octahedra in Fig.~\ref{fig:Co_BIF}), which is layered along the crystallographic c axis in the bulk.  In these BIFs, as illustrated for  Co-BIF, metal ions are found in an octahedral environment, as the metal binds to a N atom on 6 imidazolate rings, as shown in Fig.~\ref{fig:Co_BIF}. 

The boron cations connect three of these metal-imidazolate complexes, as seen in Fig.~\ref{fig:Cu_BIF}. Because of this, the exchange path between metal centers is through not only one imidazolate ring, as in ZIFs, but follows an extended Me-Im-B-Im-Me path, in both cage and 2D BIFs. This extended exchange path leads to low magnetic exchange between magnetic metal centers, if any at all . While the cage Cu-BIF MOF shows magnetic interactions close to zero, weak magnetic interactions have been detected in 2D layered BIFs~\cite{Davis2023}.

%\begin{figure}
 %   \includegraphics[width=.8\linewidth]{Co_BIF_layer.PNG}
 %   \caption{Co-BIF layer, viewed along the out-of-plane direction. Co: purple, C: brown, N: gray, O: red, B: green. H atoms omitted for clarity.}
 %   \label{fig:Co_layer}
%\end{figure}
%\begin{figure}
%    \includegraphics[width=.8\linewidth]{Co_envt.png}
%    \caption{Octahedral Co environment in Co-BIF. Co: purple, C: brown, N: gray, H: pink.}
%    \label{fig:Co}
%\end{figure}

We show that we can identify the linker vibrations that are independent of the metal ions and their environment, while the vibrations of the metal environment itself are sensitive to the substitution of a metal atom. 
Despite the differences in crystal structures, the Raman vibrations of the imidazolate linkers observed above 800~\cm~are similar for these MOFs. In contrast, the lattice vibrations observed below 100~\cm~are fingerprints of the structure. The intermediate frequencies in the range of 100-300~\cm~belong to the Raman vibrations of the metal environment and depend on the metal cation. These results demonstrate the distinct energy scales of  vibrations of MOFs of different origins. This fact allows for an easy interpretation of the vibrational features of these compounds.   

\section{Experimental}
MOFs were synthesized following reported literature procedures for Zn-BIF~\cite{Wen2017}, Cu-BIF~\cite{Banerjee2022a}, and Co and Ni-BIF~\cite{Banerjee2022}. 

Raman spectra were measured using a micro-Raman option of a T64000 Horiba-Jobin-Yvon spectrometer equipped with an Olympus microscope and a LN$_2$ cooled CCD. Spectra were excited with the 514.5 nm line of a Coherent Innova 70C laser, with the power kept below 500 $\upmu$W for a probe of 2 $\upmu$m in diameter to avoid heating the sample. Spectra were measured at room temperature with a spectral resolution of 2 cm$^{-1}$ (low-energy region) and 6 cm$^{-1}$ (Ni, Zn, and Cu in high energy region). 

Intensity of Raman spectra for different materials were normalized on the laser power, grating reflectance, and size of the slit. Background signals of stray laser light were subtracted manually from the low-frequency Co, Cu, and Ni-BIF spectra to highlight the narrow vibrational mode peaks. A Lorentzian peak centered at 0 cm$^{-1}$ was used to approximately model the background to be subtracted. In Zn-BIF, a broad photoluminescent background was subtracted by comparison to the spectra of pure imidazolate containing an identical background. To compare Raman scattering intensities of measured BIFs vibrational spectra shown in the figures are additionally  normalized to  the $\nu$(CN) modes in the 1200-1250 cm$^{-1}$ range, as a result intensity of Raman spectra $I(\omega)$ were multiplied by a constant $c_1$, where $c_1$ = 6.5 for Zn-BIF, $c_1$ = 16 for  Co-BIF, and $c_1$ = 10 for Cu-BIF. Similarly, low-frequency spectra were multiplied by a constant $c_2$ to highlight lattice modes below 100~\cm, where $c_2$ = 2.5 for Zn-BIF, $c_2$ = 6.5 for Co-BIF, and $c_2$ = 10 for Cu-BIF.

\section{Results and Discussion}

\begin{figure}
    \includegraphics[width=\linewidth]{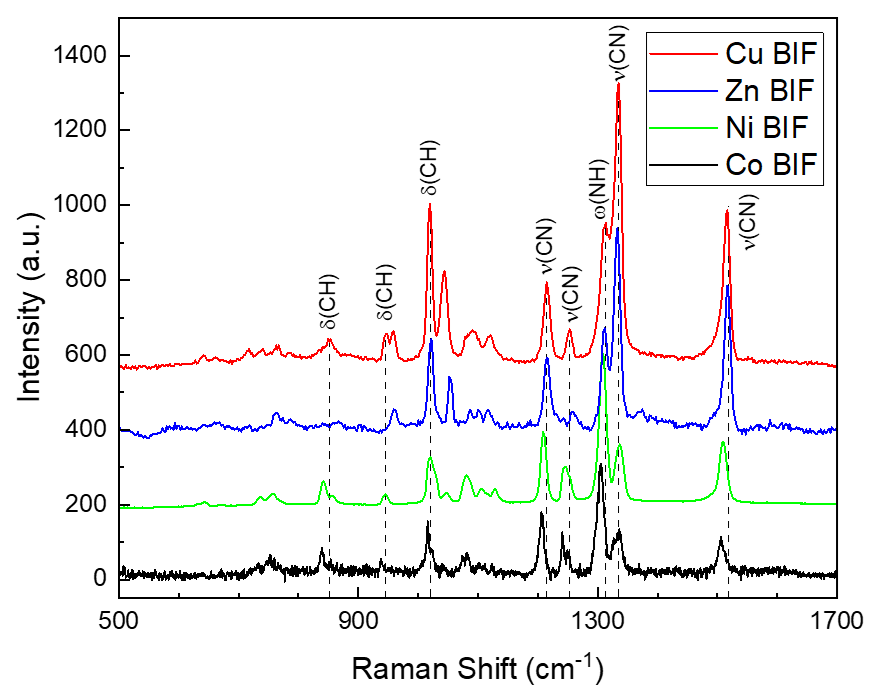}
    \caption{Raman spectra of Cu-, Co-, Zn-, and Ni-based BIFs in the range of ligand vibrational modes.}
    \label{fig:BIF_spectra}
\end{figure}

\begin{figure}
    \centering
    \includegraphics[width=\linewidth]{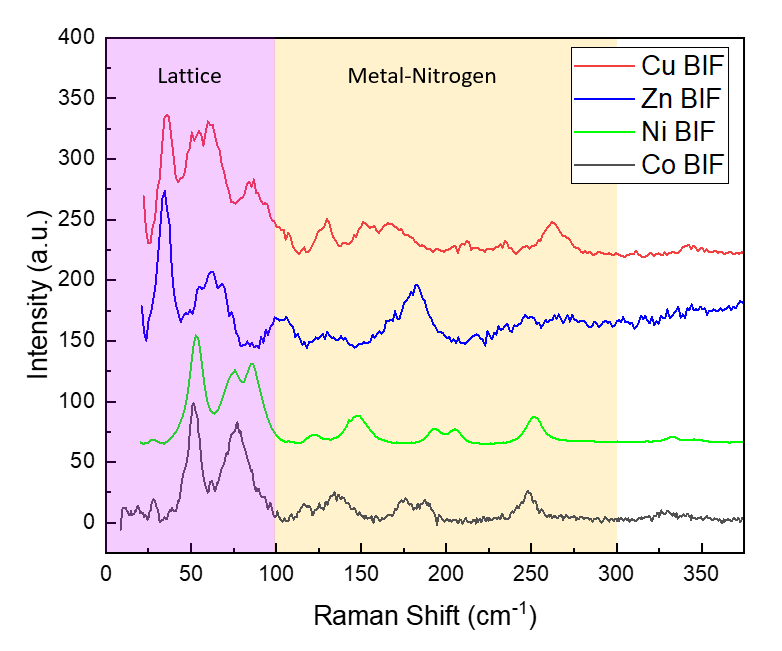}
    \caption{Raman spectra of Cu-, Co-, Zn-, and Ni-based BIFs in the low energy region.}
    \label{fig:BIF_spectra_triple}
\end{figure}

Measured Raman spectra of Co-BIF, Cu-BIF, Ni-BIF, and Zn-BIF are shown in two different regions in Figs.~\ref{fig:BIF_spectra} and \ref{fig:BIF_spectra_triple}. The 500-1700 cm$^{-1}$ region in Fig.~\ref{fig:BIF_spectra} contains the expected vibrational modes of imidazolate and allows for preliminary band assignments, which are summarized in Table~\ref{ZIFvsBIF}. The 0-500 cm$^{-1}$ region in Fig.~\ref{fig:BIF_spectra_triple} contains low-energy lattice modes below 100 cm$^{-1}$ and vibrational modes of metal environments above.

The Raman vibrations of imidazolate linkers are expected in the 800-1600 cm$^{-1}$ spectral range based on assigned Raman spectra of ZIF-8~\cite{Hadjiivanov2021}. Raman spectra of the four different BIFs have very similar frequencies of the majority of the molecular vibrations of the imidazolate ligand, despite the fact that they have three different crystal structures and four different metal nodes. This is in agreement with the expectation that the internal structure of the ligand remains unchanged between these MOFs. A comparison to previously reported Raman scattering spectra of ZIF-8~\cite{Hadjiivanov2021} reveals a direct correspondence between molecular vibrations of imidazolate in ZIF and BIF structures and is taken as a basis for our interpretation of the vibrational spectra of BIFs. Table~\ref{ZIFvsBIF} summarizes the frequencies of 8 molecular vibrations, which are consistent with previously reported band assignments for imidazolate ligand. We mark these assignments in the  Raman spectra of the BIFs  shown in Fig.~\ref{fig:BIF_spectra}.

The most intense modes in the spectral region  of 800-1000 cm$^{-1}$ are  bending vibrations of C-H bonds, which are out-of-plane with respect to the plane of the imidazolate ring. These three bands are found at very similar frequencies for all measured BIFs and close to the reported frequencies in ZIF-8. The C-H bonds are the farthest away from the metals, so one would expect that their frequencies are not strongly dependent on the metal nodes.

The spectral range between  1000 and 1500 cm$^{-1}$ contains 4 stretching vibrations of C-N bonds as well as a wagging vibration of N-H. Though the positions of the two lowest bands are shifted by ~70 cm$^{-1}$ to higher frequencies compared with the ZIF-8 spectra, the spacing of ~40 cm$^{-1}$ between them and their relative intensities are consistent with reported ZIF spectra. The C-N stretching vibration in the range from 1334-1343 cm$^{-1}$ displays a shift to lower frequencies   of ~50 cm$^{-1}$. These differences between ZIF and BIF C-N vibrations can be attributed to the fact that the two N on an imidazolate ring bind to one metal ion and one B, rather than two Zn ions as in ZIFs. Hence, the molecular vibrations that involve the N are expected to differ qualitatively between ZIF and BIF spectra, though not between BIFs themselves. In contrast, other molecular vibrations (such as C-H bending) display more consistency between ZIF and BIF spectra due to their relative isolation from the environment external to the imidazolate ring.

The reported bending vibration of the methyl group in ZIF-8 is absent in BIF spectra since  the imidazolate rings in the BIFs have no capping methyl group. The total suppression of the strong imidazolate ring breathing mode, found at 683~\cm~ in ZIF-8, can be a result of the lower symmetry (the absence of mirror symmetry) in the linker in BIF structures, which comes from imidazolate binding to one B and one metal ion, rather than two metal ions as in ZIFs. Alternatively, the suppression of this mode could be a result of a reduction of electronic density on the imidazolate ring.

\begin{table}
\caption{\label{ZIFvsBIF}Frequencies of Raman active vibrations of ZIF-8~\cite{Hadjiivanov2021} linkers compared to that of Co, Cu, Zn, and Ni-BIF. Assignment of the modes is based on the assignment of the Raman spectra of ZIF-8 ~\cite{Hadjiivanov2021}. $\delta$: bend, $\nu$: stretch, $\omega$: wag. All frequencies are listed in cm$^{-1}$. $\delta$(CH) are out-of-plane with respect to the imidazolate plane.}
\begin{ruledtabular}
\begin{tabular}{ccccccc} 
\hline
Raman Mode & ZIF-8   & Co-BIF & Cu-BIF & Zn-BIF & Ni-BIF\\ 
\hline
$\delta$(CH) & 834 & 831 & 853 & weak & 841\\ 
\hline
$\delta$(CH) (C4-C5) & 945 & 939 & 946 & 960 & 945\\
\hline
$\delta$(CH) (C2) & 1023 & 1018 & 1020 & 1021 & 1021\\
\hline
$\nu$(C5-N) & 1143 & 1214 & 1215 & 1216 & 1210\\
\hline
$\nu$(C-N) & 1182 & 1258 & 1252 & 1259 & 1247\\
\hline
$\omega$(N-H) & 1312 & 1316 & 1314 & 1311 & 1309\\
\hline
$\nu$(C5-N) & 1385 & 1343 & 1335 & 1333 & 1336\\
\hline
$\nu$(C2-N) & 1507 & 1505 & 1515 & 1517 & 1510\\
\hline
\end{tabular}
\end{ruledtabular}
\end{table}

The spectral range of  100-300~\cm~is the region of metal-ligand vibrations. The frequencies of these vibrations should be dependent both on the metal and on its coordination~\cite{Andersson2010}. In particular, literature data suggest stretching vibrations of Cu-N in the octahedral environment at 280-290~\cm, and Zn in a tetrahedral environment at 207~\cm. \cite{Andersson2010} Modes in this region therefore belong to the N-metal vibration, and have frequencies which depend on the metal ion. Zn-BIF has one mode at 181~\cm, while Co, Cu, and Ni have four weak modes at frequencies summarized in Table~\ref{MetalNvibrations}. The similarity between Ni and Co-BIF N-metal vibrations is consistent with the identical octahedral environments or Ni and Co, with the small shift in frequencies potentially arising from a change in electronic density on the N atoms in the metal environment. The more significant differences in this region in the Cu and Zn-BIF spectra are consistent with their unique metal environments.

\begin{table}[ht!]
\centering
\begin{tabular}{||c c c||} 
\hline
Co-BIF & Cu-BIF & Ni-BIF\\ 

\hline
136 & 129 & 148\\
\hline
175 & 151 & 193\\
\hline
188 & 169 & 205\\
\hline
248 & 263 & 252\\
\hline
\end{tabular}
\caption{Frequencies for four metal-N vibrational modes in the 100-300~\cm~region. Zn-BIF not included due to the lack of four clear mode in this region. All frequencies given in~\cm.}
\label{MetalNvibrations}
\end{table}

Fig.~\ref{fig:BIF_spectra_triple} presents the low frequency spectra of the BIF MOFs. Typically for molecular crystals the region below 100 cm$^{-1}$ corresponds to lattice vibrations, which would depend on the structure of the MOFs. While to the best of our knowledge, there is no spectroscopic information on the lattice modes of BIFs, the lattice ``collective'' modes of ZIFs were studied to some extent by THz spectroscopy, neutron scattering, and DFT calculations, and were also found below 100~\cm.\cite{Ryder2014,Moslein2022} 

In our experimental data, we find that two low-frequency vibrations of the cage Cu and Zn-BIFs occur at similar frequencies, while 2D Ni and Co-BIFs have similar low-frequency spectra, distinct from the cage BIFs (see  Table~\ref{BIFlatticemodes}). This demonstrates a distinct dependence of the lattice mode spectra on the structure of the MOFs, showing that the lattice vibrations are  fingerprints of a particular lattice structure, and not the chemical environment.  We find that two of the lattice modes observed in the spectra for the two cage BIFs, Cu and Zn, are similar to Raman-active THz modes in the calculated DFT spectra of ZIF-8, which is also composed of a porous cage-like structure. These include a strong mode at 33.36 cm$^{-1}$, assigned to a symmetric 4-membered ring (see Fig.~\ref{fig:Cu_BIF},\ref{fig:Zn_BIF}) gate opening, and a mode at 64.61 cm$^{-1}$, assigned to a 4-membered ring shearing~\cite{Ryder2014} .

\begin{table}[ht!]
\centering
\begin{tabular}{||c c c c c||} 
\hline
ZIF-8 DFT (\cm) & Zn-BIF & Cu-BIF & Co-BIF & Ni-BIF\\ 
\hline
ring gate opening (33.36) & 34 & 35 & \textbf{---} & \textbf{---}\\
\hline
\textbf{---} & \textbf{---} & \textbf{---}& 52 & 53\\
\hline
 ring shearing (64.61) & 62 & 58 & \textbf{---} & \textbf{---}\\
\hline
\textbf{---} & \textbf{---} & \textbf{---} & 77 & 74\\
\hline
\end{tabular}
\caption{Frequencies of observed modes below 100 cm$^{-1}$ for the four BIFs.}
\label{BIFlatticemodes}
\end{table}

The 2D layered BIFs, on the other hand, do not display any similarity to calculated THz modes of ZIFs, which is consistent with the significant structural differences between the 2D triangular lattice structure and the various cage structures. DFT calculations on the 2D structures are necessary to further interpret these vibrations.

\section{Conclusions}

In this work we have presented Raman scattering spectra of BIF MOFs with different metal ions in a frequency range between 30 and 2000~\cm\, and demonstrated a separation between the energy scales related to the different types of vibrations of the MOFs. The high frequency region of the spectra contains Raman-active vibrational modes of imidazolate ligands, which are similar between four different BIFs with different structures and metal nodes. They can be assigned based on the literature data of MOFs with the same linker molecules, ZIFs.

The low-energy region of the spectra demonstrates a consistency in lattice modes of isostructural BIFs. The modes of the cage BIFs have frequencies similar to the calculated lattice modes for ZIFs, while 2D BIFs have a distinctly different lattice mode spectrum.  These results demonstrate that the lattice vibrations of MOFs are fingerprint of a certain structure. 

The spectral region between approximately 100 and 300 cm$^{-1}$ is dominated by the vibrations of the metal environment. The frequencies of vibrations observed in this region depend on the metal atom, but show some consistency between BIFs with different metals in the same environment.

The consistency of lattice modes and ligand vibrations provides an efficient way to understand and assign the spectra of MOFs without performing full calculations of the vibrational response of each MOF structure.

\begin{acknowledgements}
Acknowledgement is made to the donors of the American Chemical Society Petroleum Research Fund for   partial support  of this research. We  acknowledge the support of  NSF Award No. DMR-2004074. S. B., P. B.-V., and V. S. T. acknowledge support by the U.S. Department of Energy (DOE), Office of Science, Office of Basic Energy Sciences, Catalysis Science program, under Award DE-SC0021955. P. B-V. thanks the Dean’s ASPIRE grant from the Office of Undergraduate Research, Scholarly and Creative Activity at Johns Hopkins University. We further acknowledge Professor Tyrel McQueen, Dr. Veronica Stewart (Department of Chemistry, Johns Hopkins University), Chris Lygouras, and Peter Orban (Department of Physics and Astronomy, Johns Hopkins University) for their assistance in obtaining magnetic susceptibility and magnetization data.
\end{acknowledgements}

\section*{Data Availability Statement}

The data that support the findings of this study are available from the corresponding author upon reasonable request.

\section*{Conflict of Interest}

The authors have no conflicts to disclose.

% If in two-column mode, this environment will change to single-column format so that long equations can be displayed. 
%\begin{widetext}
%$$\mbox{put long equation here}$$
%\end{widetext}

% Figures should be put into the text as floats. 
% Use the graphics or graphicx packages (distributed with LaTeX2e).
% Put the label that you will use with \ref{} command in the braces of the \label{} command.
%
% \begin{figure}
% \end{figure}

% \begin{table}
% \end{tabular}
% \end{table}

%\begin{acknowledgments}
% Put your acknowledgments here.
%\end{acknowledgments}

\nocite{*}
\bibliography{BIF_Raman}
%\printbibliography
\end{document}